\documentclass[conference]{IEEEtran}
\IEEEoverridecommandlockouts
\usepackage{cite}
\usepackage{amsmath,amssymb,amsfonts}
\usepackage{algorithmic}
\usepackage{subcaption}
\usepackage{graphicx}
\usepackage{hyperref}
\usepackage{textcomp}
\usepackage{xurl}
\usepackage{xcolor}

\def\BibTeX{{\rm B\kern-.05em{\sc i\kern-.025em b}\kern-.08em
    T\kern-.1667em\lower.7ex\hbox{E}\kern-.125emX}}

\begin{document}

\title{Quantum computing online workshops and hackathon for Spanish speakers: A case study
}

\author{\IEEEauthorblockN{Alberto Maldonado-Romo}
\IEEEauthorblockA{\textit{Centro de Investigación en Computación} \\
\textit{Instituto Politécnico Nacional}\\
Mexico City, Mexico \\
alberto.maldo1312@gmail.com}
\and
\IEEEauthorblockN{Lia Yeh}
\IEEEauthorblockA{\textit{Department of Computer Science} \\
\textit{University of Oxford}\\
Oxford, UK \\
lia@fullstackquantumcomputation.tech}
}


\IEEEpubid{\begin{minipage}{\textwidth}\ \\[12pt]\\
\copyright2022 IEEE.  Personal use of this material is permitted.  Permission from IEEE must be obtained for all other uses, in any current or future media, including reprinting/republishing this material for advertising or promotional purposes, creating new collective works, for resale or redistribution to servers or lists, or reuse of any copyrighted component of this work in other works.\hfill
\end{minipage}} 

\maketitle

\begin{abstract}
We discuss the challenges and findings of organizing an online event in Spanish, consisting of a series of introductory workshops leading up to a quantum hackathon for Latin America.  220 Spanish speakers were registered, 66\% of whom self-identified as being at an introductory level of quantum computing.  We gain a better picture of the impact of quantum computing in Latin America, and the importance of generating educational resources in Spanish about quantum computing.  Additionally, we report results on surveying the participants by country; educational status; self-reported levels of quantum computing, linear algebra, and Python competency; and their areas of interest within quantum.

This event was organized by Quantum Universal Education with the Centro de Investigación en Computación del Instituto Politécnico Nacional (CIC-IPN) as the host institution, in collaboration with a number of organizations and companies: IBM Quantum, Xanadu, Multiverse Computing, Quantum Universal Education, Quantum Hispano, QMexico, Haq.ai, Dive in Learning.
This was part of a larger event, the Qiskit Fall Fest 2021, as one of several hackathons organized around the world in a similar span of time.  In each Qiskit Fall Fest hackathon, participants were challenged to form teams of up to 5, to develop in 5 days a project using the IBM Qiskit framework.
\end{abstract}


\section{Introduction}
Accompanying the rise in applications of quantum technology, there has been growing interest in quantum education and workforce development initiatives~\cite{b1}.  These beginning and/or grassroots efforts have the potential to be very impactful, particularly in this formative period, where traditional educational offerings have yet to catch up to train the skills demanded for in these new industries~\cite{b2}.  As with any emerging technical subject, most students and professionals across all ages and stages of learning who would like to begin to learn it, will not find a course offering at their institution to gain exposure to the subject.  This makes it important that there be informal educational opportunities open to those who would like to learn, and welcoming to those new to the topic, especially as one must first be exposed to the topic in order to then become interested in learning it.  Much progress has been made on addressing these needs in recent years, from online extracurricular courses for middle and high school students~\cite{b3} to games and software tools designed to teach quantum science concepts~\cite{b4}.  This report serves to introduce the approach of hackathons as a gateway for quantum computing learning and outreach.

Hackathons are events where participants form teams of one to up to around five participants, who compete to create a functioning project from start to finish, over the course of the event, which can range from one day to a few weeks in duration.
Hackathons are a widely adopted approach to facilitating learning and collaboration in a short but focused time period, with hundreds of hackathons taking place each year.  Definitions of hackathons, or ``hacking marathons,'' may vary greatly depending on whether the intended purpose(s) is innovation, collaboration, competition, business solutions, software prototyping, education, outreach, research, or fun~\cite{b4,b5}.

Despite the growing number of quantum hackathons worldwide, there is limited documentation and evaluation of their organization and effectiveness.
A particular subcategory of quantum hackathons has been studied for educational purposes: quantum games hackathons, and their more relaxed and less competitive counterpart, quantum game jams.  Ref.~\cite{b4} presented a comprehensive overview of quantum games --- games with quantum elements in their game mechanics --- for education and outreach, with a section describing an online Quantum Games Hackathon.  Ref.~\cite{b6} evaluated each quantum game created in the course of five quantum game jams on criteria such as playability and educational value.

\IEEEpubidadjcol
\section{A Spanish-language virtual hackathon for Latin America}
As an emerging area worldwide, quantum computing has seen recent growth in the educational materials accessible in a variety of forms, including but not limited to: formal courses, textbooks, self-paced online courses, video series, programming tutorials, workshops, games, and comics.  However, all these learning resources and opportunities are predominantly restricted to the English language.
For one to become interested in quantum and feel a sense of belonging, quantum education needs to recognize, and strive to attune to, each and every diversity of identity and background.
In order to drive equitable and inclusive education in quantum science and technology, the challenges of language, geographical, and socioeconomic barriers must be addressed.

In Mexico, the Cuarantena Cuantica (literal translation: Quarantine Quantum) quantum computing seminar~\cite{b7} was held virtually in January 2021 by the Sociedad Científica Juvenil (SCJ), i.e. the Juvenile Scientific Society of Mexico.  It was followed by the Qiskit Summer Jam 2021 Mexico quantum hackathon~\cite{b8} held in August 2021 as a collaboration between SCJ and QMexico, a QCousins branch of the QWorld non-profit organization.  These and the Quantum Latino 2021 event~\cite{b9} were the direct antecedents of the workshop series and hackathon described in this report, as quantum computing events respectively centered on Mexico and Latin America.

With the importance of supporting more accessible pathways to exposure to quantum concepts in a collaborative and welcoming environment in mind, this event was organized with beginner friendliness as the key priority.
The goal was to encourage participation at all levels, especially for those for whom this was their first quantum event or even exposure to technical quantum concepts and quantum programming.
For this reason, we put together a series of workshops in Spanish and English in the week preceding the hackathon.  By design, the first workshop, offered once in Spanish and once in English by the Quantum Universal Education not-for-profit organization, was Introduction to Quantum Computing; it began by explaining what a qubit is logically and what it can look like physically, with the quantum circuits introduced alongside small illustrations of colorful cats, accompanied by a live code demonstration in the quantum programming language Qiskit.
As we conclude in the section on our participant survey findings, this decision of focusing on introductory workshops was crucial to the objective of increasing awareness of quantum computing in Latin America.

With these goals in mind, we proceed to describe the organizational structure, dissemination of event sign-up information, and other considerations for the planning of the event, in hopes of informing future Spanish language and/or hackathon quantum educational events.

\subsection{Overview of the event organization}

The hackathon was part of IBM Quantum Education's Qiskit Fall Fest 2021, which coordinated the initiative to encourage and support 18 hackathons organized in roughly the same time frame around the world.
IBM Quantum Education provided guidance in planning a hackathon in the form of courses and a hackathon guide~\cite{b10}.  These courses supported the planning of this hackathon to design it online, and to make it open to all Spanish speakers.
Topics covered included the dissemination of information, the tools to broadcast, and platforms to manage and link people together whether that be in-person or online. Some tools were Hype Innovation Management Software, a web platform used to manage, for each hackathon: team formation, project ideation, and project submission; and Discord, a community and messaging social media platform used by each hackathon for schedule, announcements, networking, mentoring, memes, and other communication. Considering the locations of countries in Latin America, to create a schedule that the most people across these time zones can realistically participate in, we used the UTC-6 time zone.

The event, while being run online, was hosted thanks to the Centro de Investigación en Computación del Instituto Politécnico Nacional (CIC-IPN).  For reasons due to the pandemic situation, the in-person conditions were not suitable.  With regards to the CIC-IPN organization, the event was part of the CORE International Congress 2022, an international congress focused on computer science, organized by students from the Centro de Investigación en Computación (CIC) in Mexico City~\cite{b11}.
In conjunction with the Instituto Politécnico Nacional (IPN), this facilitated extending the scope of participation to different institutions in the city, country, and different countries.

\begin{figure}[htbp]
\centerline{\includegraphics[width=9cm]{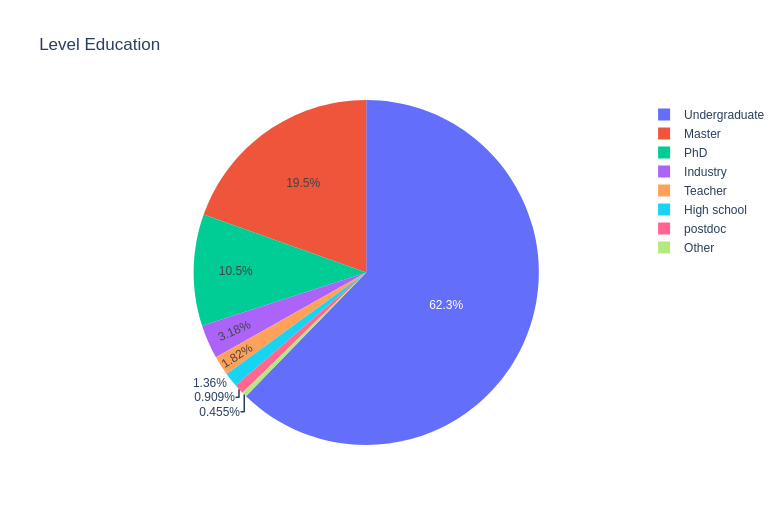}}
\caption{Percentage of registered attendees by education level.}
\label{social_medias}
\end{figure}

To reach Spanish speakers online, information on the event and how to register were disseminated through a number of online channels, as a collaboration between various organizations and communities each with hundreds or thousands of people involved.  Communities of special mention for the unique roles each played in supporting this event include Quantum Universal Education, Quantum Hispano, the QMexico chapter of the QWorld non-profit, and the Qiskit Advocates program of IBM Quantum Community.  The social networks and messaging platforms utilized included Facebook, Twitter, LinkedIn, Discord, Whatsapp, Slack and others, with a majority of registrants having heard of the event through Facebook (see Figure \ref{social_medias}).  The announcement campaign was conducted in both English and Spanish, using phrases such as introduction to quantum computing; programming quantum algorithms; quantum computing hackathon; and create and learn your first quantum computing project.
For there to be an expectation as to the prerequisite knowledge recommended, the suggested skills to have an introductory grasp of were: how to program in Python, to know linear algebra, and to know probability; to know about quantum mechanics or quantum computation was optional, given the purpose of the workshops.


The event was scheduled such that the workshops took place across six consecutive days.
The last day of workshops focused on industry, where presenters from Multiverse Computing and Xanadu offered workshops by Spanish speakers where they indicated the aspects and requirements of a job in a quantum computing company, and different tips for starting out in this area.
The week of workshops was followed by the hackathon, which took place across five consecutive days.
Over the course of the hackathon, people new to quantum would have ample time to get to know each other and get started in the creation of their first quantum computing project with virtual sessions, and not feel rushed to ask general questions and concerns about quantum computing or about their projects.

\pagebreak

\subsection{During the Event}

The event consisted of two parts in order to involve people who had no knowledge of quantum computing, but had some general knowledge of Python programming, linear algebra, and optionally general topics of quantum physics. For the event, a server was created on the Discord social community platform, in which only registered participants had access to the join link.  After entering the Discord server, participants could select the option to attend the workshops, where they were then automatically provided access to the channels where the links, resources, and question and answer sessions regarding each workshop were posted.  Likewise, participants could select the option to attend the hackathon, to access channels where they can introduce themselves, describe their areas of interest, form teams, and discuss possible projects and ideas.  For each team to come up with their project proposal, they are given a channel for their team to work, visible to just their team members and to the hackathon volunteers and mentors to respond to any questions and concerns.  

\begin{figure}[htbp]
\centerline{\includegraphics[width=9cm]{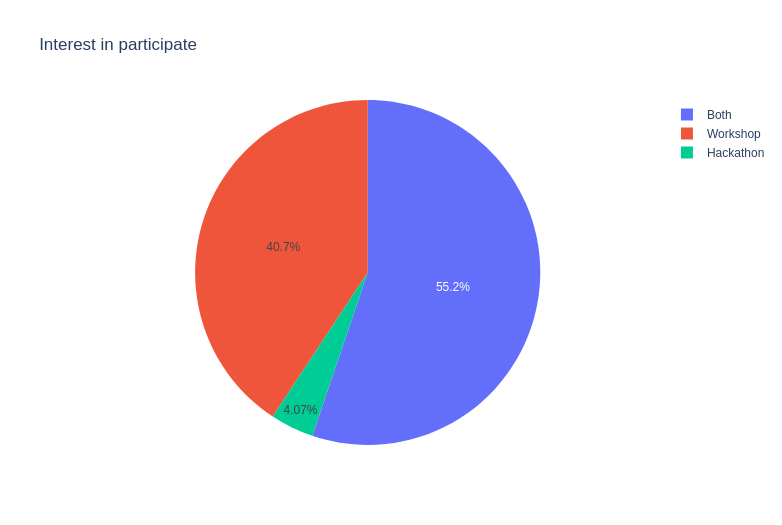}}
\caption{Percentage of registered attendees interested only in the hackathon (4.07\%), only in the workshops (40.7\%), and in both the workshops and the hackathon (55.2\%).}
\label{worshop_hackathon}
\end{figure}

For all participants, whether they opted to take part in the workshops, hackathon, or both (See Figure \ref{worshop_hackathon} for these choices at the time of registration), there were general channels for: announcements about quantum computing news, sharing how to get started with tutorials and applications, suggesting project ideas, and memeing.
These were used as spaces to motivate and support people to get comfortable, challenge themselves, learn from and with each other, and let their imagination take off.
They were encouraged to ask or inquire more about topics in quantum, and to take a look at the linked open educational resources accessible to them geared towards those new to quantum concepts, for example from IBM Quantum Education and Quantum Universal Education.

\subsection{Series of Workshops}

The series of workshops was designed to introduce at a beginner level different applications of quantum computing, such as optimization, chemistry, algorithms, machine learning, and video games.  These are areas focused on in previous Qiskit hackathons, as areas for which there are a variety of applications and more accessible learning materials.
In addition to technical workshops, a panel was added to shed light on the landscape of quantum computing in Latin America, and opportunities for work and perspectives shared by Spanish-speaking people who are in the industry.
The purpose of these workshops is to introduce the state of the art in quantum computing to Spanish speaking audiences, for which the original materials have to be adapted to Spanish and adapted to an introductory level. For this we considered the criterion of inviting people who meet one of three requirements: 1) be a Qiskit advocate with a particular interest in an area of quantum computing; 2) be someone who works in the quantum computing industry being people from for instance IBM Quantum, Xanadu and Multiverse Computing; or 3) have completed an internship or mentorship in one of the quantum computing communities such as QWorld.  Each of the selected workshop speakers facilitated a workshop in their area of expertise.  For accessibility of the learning in general, the workshops were coordinated to be introductory and limit the need for prerequisite knowledge of quantum where possible.  Moreover, the more introductory topics were scheduled earlier in the week, and the programming examples were encouraged to use the Qiskit framework to ensure continuity of understanding between workshops.
The series of workshops and their descriptions can be seen in Table 1.

\begin{table}[htbp]
\caption{Series of Workshops}
\begin{center}
\begin{tabular}{|p{8cm}|}
\hline \\
\textbf{Introduction to quantum computing}\\
Language: English / Spanish\\
Workshop presenter: Lia Yeh / Alberto Maldonado Romo\\
Description: There is a growing trend towards quantum computing. Companies and governments are investing in its research and development since its use represents the answer to tasks that involve excessively complex calculations in artificial intelligence, chemistry, cryptography, among others. Actors such as IBM and Microsoft are looking to train developers who can program their quantum computers. To this end, they make their platforms freely available to us. The challenge now is to know the resources, materials, and events that one can develop all their skills.  This presentation will address the opportunity to have a vision of this large and complex world.\\
\\ \hline \\

\textbf{Estado actual y perspectiva de los juegos cuanticos}\\
Language: Spanish\\
Workshop presenter: Anamar\'ia Garc\'ia Hern\'andez\\
Description: The field of quantum games is developing rapidly. In this project we collect information about quantum games at present and classify them into different categories. The results obtained will be discussed, as well as future projections.\\
\\ \hline \\

\textbf{Creacion de oracles para algoritmos}\\
Language: Spanish\\
Workshop presenter: Emilio Pelaez\\
Description: In this workshop, we will explore the construction of oracles for different algorithms. Giving examples on concrete algorithms such as Grover's search algorithm and the 3-SAT problem, and defining the concept of an oracle formally and how we can translate it into a circuit efficiently.\\
\\ \hline \\

\textbf{Calculando observables f\'isicos con VQE}\\
Language: Spanish\\
Workshop presenter: Siddhartha Morales\\
In this workshop, we will explore how to use variational algorithms to solve some interesting physical problems, such as the ground state of a molecule and of an atomic nucleus. We will see how to use the quantum variational algorithm, as well as how to create our own ansatz, test different optimizers and how to send the work to a real quantum computer.\\
\\ \hline \\

\textbf{Algoritmo BB84}\\
Language: Spanish\\
Workshop presenter: Luis Mart\'inez\\
Description: In this workshop, we will see a brief introduction to the area of quantum cryptography. With special emphasis on the BB84 key exchange algorithm. We will also review some primitives that are important in this field. For this workshop it is not necessary to know classical cryptography as the basic concepts will be reviewed in the workshop.\\
\\ \hline \\

\textbf{Knapsack Problem}\\
Language: Spanish\\
Workshop presenter: Claudia Zendejas Morales\\
Having quantitative information to make decisions leads to more and better profits. Studying the knapsack problem (KP) allows us to find solutions to combinatorial optimization problems by modelling a situation analogous to filling a knapsack. Its applications range from transportation and logistics problems to financial investments. In this workshop, we will see how to solve the knapsack problem with quantum computing.\\
\\ \hline
\end{tabular}
\label{tab1}
\end{center}
\end{table}

\begin{table}

\begin{center}

\begin{tabular}{|p{8cm}|}
\hline
\textbf{Criptograf\'ia muri\'o RSA?}\\
Language: Spanish\\
Workshop presenter: Daniel Sierra-Sosa\\
Description: One of the areas of interest in Quantum Computing is information security. It involves the fundamental principles and techniques of quantum computing, notions of information theory, algorithms (Grover's search and Shor's factorization), and applications of Quantum Computing such as quantum encryption and key distribution. This workshop will aim to explore and discuss real-world scenarios related to information security in times of quantum technologies, participants will understand the opportunities and challenges in this area and will have a hands-on experience on the IBM Q Experience platform.\\
\\ \hline \\

\textbf{Quantum Game Development at an introductory level}\\
Language: English\\
Workshop presenter: Wen-Sen Lu\\
Description: It is our privilege to explore the cutting-edge quantum computational space during the NISQ era with QISKit. Looking back into the history, especially in the 1970’s, arcade game developers already started the machine-level programming and prepared themselves as the future coders even if the hardware was still limited. In the meanwhile, game-driven breakthrough for the classical hardware, such as the first 3D acceleration chip Super FX in Nintendo super-NES home console, also demonstrated the possibilities where new hardware could be inspired by the game developers. In this talk, I will start with my personal experience to quickly walk us through the process of quantum game development: looking for ideas from existing games, selecting development tools, and putting together two example codes in PICO-8 (Lua) to quickly demonstrate the classical and quantum counterpart of the game dev, respectively.\\
\\ \hline \\

\textbf{Haq.ai}\\
Language: English\\
Workshop presenter: Adam Fattal\\
Description: haq.ai is a platform oriented for everyone interested in the field of quantum computing that wants to sharpen their quantum programming skills. Through a wide collection of problems in many topics and with different difficulties, users can develop their abilities. They can learn to efficiently decompose quantum circuits through methods presented in literature, harness the power of numerical computation in the field of quantum information, explore fun problems that require using popular algorithms and protocols, and much more. We aim to make the journey through quantum computation more interactive, while we don’t offer a whole educational component, we offer a great supplement to a conventional quantum computing education.\\
\\ \hline \\

\textbf{Introducci\'on al Aprendizaje de Máquina Cu\'antico}\\
Language: Spanish\\
Workshop presenter: Alberto Maldonado Romo\\
Description: In this workshop we will see a small introduction of how to pass classical information to qubits to be able to treat them and get to general models such as neural networks in their quantum version.\\
\\ \hline \\

\textbf{Situaci\'on de la computaci\'on cu\'antica en Am\'erica L\'atina}\\
Language: Spanish\\
Workshop presenters: Jazmin Esteva, Bruno Ram\'irez, Dr. Javier Orduz\\
Description: Quantum computing has had a great impact on the world and many companies have focused on this area, but what is happening in Latin America?\\
\\ \hline \\

\textbf{Introducci\'on al QAOA}\\
Language: Spanish\\
Workshop presenter: Victor Onofre\\
Description: QAOA is one of the hybrid quantum-classical algorithms that have been proposed to take full advantage of current quantum resources. In this workshop, we will explain its application to the MAX-CUT problem.\\
\\ \hline 

\end{tabular}
\label{tab1}
\end{center}
\end{table}

\begin{table}

\begin{center}

\begin{tabular}{|p{8cm}|}
\hline \\
\textbf{Quantum Enhanced Monte Carlo Simulations}\\
Language: Spanish\\
Workshop presenter: Cristina Sanz Fern\'andez\\
Description: Case study of how quantum computing can be useful to us today. Monte Carlo simulations are a widely used tool both in research (physics, chemistry, etc.) and in practical applications in our day-to-day life (finance, meteorology, telecommunications, etc.). In this talk, I specifically explain how quantum leads to a quadratic improvement of Monte Carlo calculations.\\
\\ \hline \\

\textbf{Carreras en la computaci\'on cu\'antica Xanadu}\\
Language: Spanish\\
Workshop presenter: Catalina Albornoz\\
Description: In this talk, we will discuss the different career opportunities in the field of quantum computing, and what skills can lead you to that dream job.\\
\\ \hline
\end{tabular}
\label{tab1}
\end{center}
\end{table}


\subsection{Hackathon}

In organizing a hackathon, a number of logistical components are involved.  In this section, we describe the aspects of team formation and project ideation; mentors; and project evaluation and prizes.  Additionally, we provide brief descriptions of and links to the winning projects.

\begin{itemize}
    \item \textbf{Team formation and choosing projects:} The workshop session included activities such as a six-phase introduction, discussion of topics of interest, and the creation of specific channels for participants to search for teammates, with teams limited to 2 to 5 members. 48 hours before the hackathon, during team formation, mentors were involved in proposing and bouncing off ideas with the participants.  This process included understanding, for each participant, what workshop they liked the most, what geographical location they are from, what skills they have and would like to learn, and their past project experiences.  Suggested project ideas and feedback on proposed ideas were made to each participating team, especially in conversation with those who had doubts as to where to start with proposing their project.
    \item \textbf{Mentors:} The mentors were selected from the Qiskit advocate program, most of them being the same people who gave workshops, or people who could speak or write in Spanish to provide one-on-one support to the teams.
    \item \textbf{Project evaluation and prizes:} The hackathon judges evaluated the projects, consisting of the project files and a short recorded talk explaining the motivation and demoing the project, according to the following four criteria, each of equal weight.
    \begin{itemize}
        \item \textit{Technical Challenge:} Did the team challenge themselves and try to learn and implement something new to them or to the area?
        \item \textit{Impact:} Does the project have a high potential for impact? E.g. industrial application, educational value, theoretical interest, etc.
        \item \textit{Creativity:} Does the project go beyond the scope of a typical hackathon project?  Unlike the technical challenge criterion, the creativity criterion incentivizes imagination in forms beyond technical (eg. artistic, originality, user friendliness, etc.).
        \item \textit{Presentation:} Is the project functioning and thorough?  Is the presentation of the project thoughtfully explained and easy to understand?
    \end{itemize}
\end{itemize}
There are some observed similarities between the above four hackathon judging criteria, and those of the quantum games hackathon by the Quantum AI Foundation of 1) Correctness, 2) Playability, 3) Originality, and 4) Quality and completeness~\cite{b4}.

To improve the learning experience, each team was asked which workshops they enjoyed, and what topics they were interested in, so that to the extent possible, they could be assigned a mentor(s) suiting those specializations.

\subsection{Winning projects}
A total of 10 projects were ideated by the 29 hackathon participants, of which 8 projects were submitted as code along with recorded presentation for judging~\cite{b12}.  Prizes were awarded to the top three projects, with an additional prize for the Best Education Hack awarded by Quantum Universal Education.

The winning projects along with brief description are in Table 2, and links to each project are in the bibliography.

\begin{table}[htbp]
\caption{Winning Projects}
\begin{center}
\begin{tabular}{|p{8cm}|}
\hline \\
\textbf{Threerra: A Qiskit module for three-level systems}~\cite{b13}\\
First Place\\
Description: Created a module to allow users to both create unitary operations acting onto three-level systems (qutrits) using Qiskit Pulse, and to execute them on real hardware available through the IBM Quantum platform.\\
\\ \hline \\

\textbf{Quantum Radio Resource Scheduling for 4G and 5G simulators}~\cite{b14}\\
Second Place\\
Description: The way to find the optimal resource allocation for users for which throughput is calculated based in resource blocks (chunks of frequency) and modulation (a sort of channel indicator).\\
\\ \hline \\

\textbf{Variational Quantum Circuits in a Protein Network Graph}~\cite{b15}\\
Third Place\\
Description: The main idea of the proposal is to map biochemical interactions inside a 3D protein structure into a graph network.\\
\\ \hline \\

\textbf{Quantum classifier for medical data}~\cite{b16}\\
Best Education Hack\\
Description: In this project, we propose an introduction to quantum machine learning using a variational quantum algorithm for classification, applying it to two medical datasets.\\
\\ \hline

\end{tabular}
\label{tab1}
\end{center}
\end{table}

Additionally, we report a follow-up to the 14 winning participants six months after the conclusion of the hackathon.

First, we remark that of the 14, 9 expressed preference for future quantum computing events to be hybrid, in comparison to 2 for in-person and 3 for virtual.

In the response to the question 
\begin{quotation}
How much time had passed between your first exposure to quantum science, computing, or engineering, and participating in this event (in October 2021)?
\end{quotation}
6 of 14 responded 0-6 months.  Despite the high percentage of beginners, in response to the question
\begin{quotation}
On a scale of 1 (not interested) to 5 (very interested), what is your interest in a career in quantum?
\end{quotation}
2 of the 14 winning participants indicated a 3, 4 indicated a 4, and 8 indicated a 5.

Most notably, 9 out of the 14 hackathon winners responded Yes to the question
\begin{quotation}
Since participating in the Qiskit Fall Fest CIC-IPN Hackathon have you participated in any quantum-related event?
\end{quotation}

This shows that the hackathon winners felt empowered to continue to actively learn quantum science following the event, which they did through participating in quantum-related school projects, challenges, summer schools, and other quantum hackathons.

\section{Survey responses of participants}

To guide inclusive and beginner-friendly hackathon organization, four questions were kept in mind~\cite{b17}:
\begin{itemize}
    \item Who is eligible to apply?
    \item Who is it marketed to?
    \item Who actually attends?
    \item Who is it prepared to support?
\end{itemize}

For participants for whom it is their first exposure to what they can do with quantum computing, it is understandable to feel not yet ready or comfortable to engage in a quantum hackathon.  Regardless of whether they chose to participate in the hackathon (which anecdotally was a very positive experience for not just those experienced in quantum, but those new to quantum as well), the organizers considered the engagement and sincerity of participation in the workshops to be just as, if not more, important than the hackathon component.

\subsection{Demographic information }

\begin{figure}[htbp]
\centerline{\includegraphics[width=9cm]{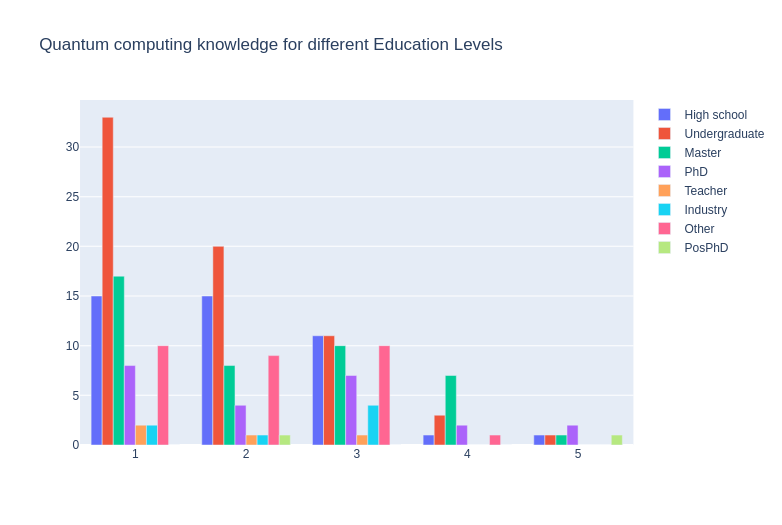}}
\caption{The survey data is first plotted by the number of participants for each self-reported level of quantum computing knowledge, ranging from 1 (little to none) to 5 (expert).  Within each of these five levels, the data was then plotted by number of participants for each educational level (if they were a student) or occupation (for non-students).  This shows that $66\%$ of the participants identified as beginners in quantum computing --- 1 or 2 out of 5.}
\label{level_int}
\end{figure}

The participation student status data indicates that the event, open to students and non-students of all levels of education and experience in quantum computing, attracted participation from all levels.  This shows that there are no noticeable gaps in the demographic reached, compared to that targeted.  We observe that amongst high school participants, the self-identified level of quantum computing was comparable to that of other participants with other levels of education and occupational status.  This raises the possibility of an alternate approach to quantum hackathons: Instead of targeting a certain level of education (e.g. high school) amongst which exposure to quantum computing may vary greatly, another option is to target beginners in quantum computing regardless of level of education, instead categorized by self-identified level of experience in quantum computing (see Figure \ref{level_int}).  This approach is informed by the student-run hackathon \textlangle{}Womxn/Hacks\textrangle{}, which in 2019 hosted $\approx 200$ female-identifying and non-gender-binary undergraduate and graduate students, $\approx 2/3$ of whom self-identified as beginners in programming, and nearing 3/4 of whom were pursuing neither computer science nor computer engineering degrees — $\approx 1/3$ of the total participants were arts and humanities students~\cite{b17}.  The primary advantage of this approach is that individuals who have had their first exposure to or gained confidence to learn the topic at a later stage in life, who are disproportionately likely to be underrepresented in that topic, can have the opportunity to learn.  For an approach like \textlangle{}Womxn/Hacks\textrangle{}’s which awarded prizes for both beginner and advanced levels, more advanced students in an earlier stage of education who may have exhausted the learning opportunities in quantum available to high schoolers have the option to seek the challenge to learn more by forming a team with others more experienced in other aspects and competing for advanced category prizes.

The country participation data indicates that the country with the most participation was Mexico.  This is as the organizers anticipated, with a number of possible explanations that this could be attributed to.  First, the advertisement for the event was distributed and shared by a number of Mexican organizations and reaching audiences of Mexican identity.  Second, while this was a fully online event, there is the status of the host university and members of the event organizers being Mexican, and hence increased technical and administrative support for those time zones.  Although the event time zones had workshops and final project presentations scheduled at more central times of day with respect to time zones around the world, the event being virtually hosted in Mexico may have resulted in people in time zones further from Mexico being less inclined to participate, due to anticipated inconvenience or impracticality of time zone difference.

\begin{figure}[htbp]
\centerline{\includegraphics[width=9cm]{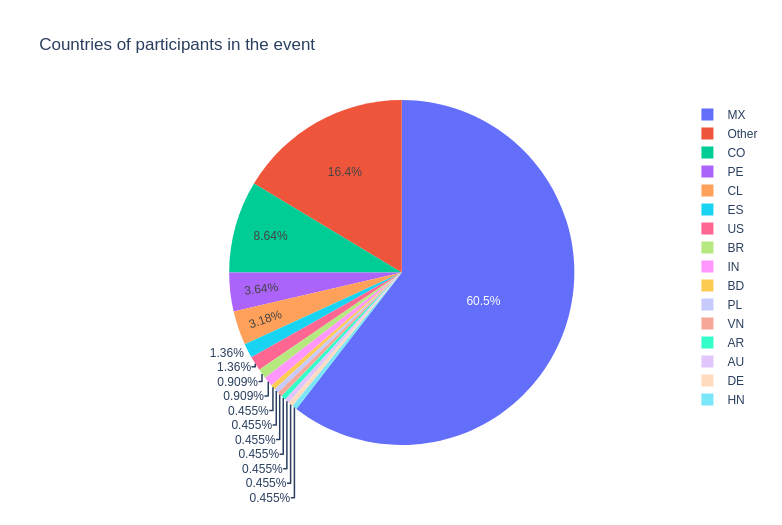}}
 \caption{Percentage of participants by country.}
 \label{country}
\end{figure}

With that said, the organizers were pleasantly surprised by the number of countries represented in the participants.  This not only indicates that educational opportunities in quantum are appealing to students of all levels and from many geographical locations, but that channels to reach out to them all exist and should be utilized more.  Combining this with the information that a majority of participants heard about the event through Facebook, this indirectly alludes to the fact that online Spanish-speaking communities for learning and interest in quantum encompass a greater diversity than previously realized (See Figure \ref{country}).

Regarding the question guiding hackathon organization, “Who is it prepared to support?”: To reach more persons in other countries, a recommendation for future online and international Spanish-language hackathons is to assemble an organizing team across more countries.  However, the amount of overhead in supporting more time zones can make organization logistically more complex, and so a balance is needed.

\subsection{Self-reported amount of experience in quantum, linear algebra, python}

\begin{figure}[htbp]
\begin{minipage}[c]{9cm}
\centerline{\includegraphics[width=9cm]{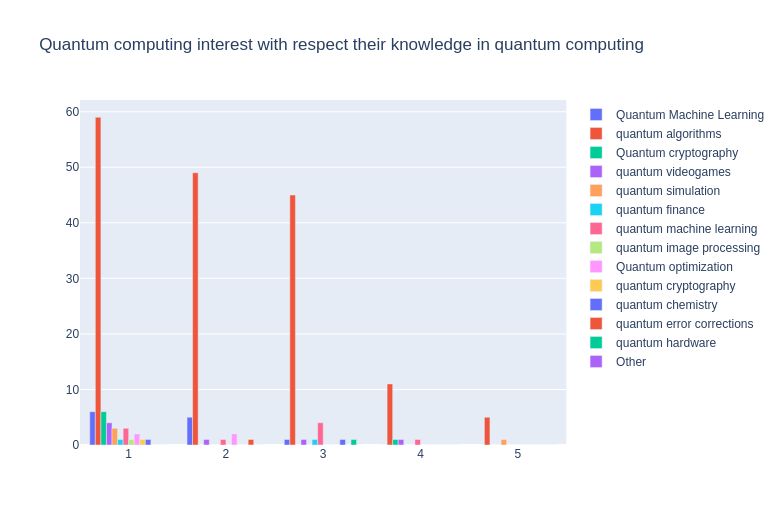}}
\subcaption{Participants' knowledge of quantum computing compared to their interest in different areas of quantum computing.}
\label{area_quantum}
\end{minipage}

\begin{minipage}[c]{9cm}
\centerline{\includegraphics[width=9cm]{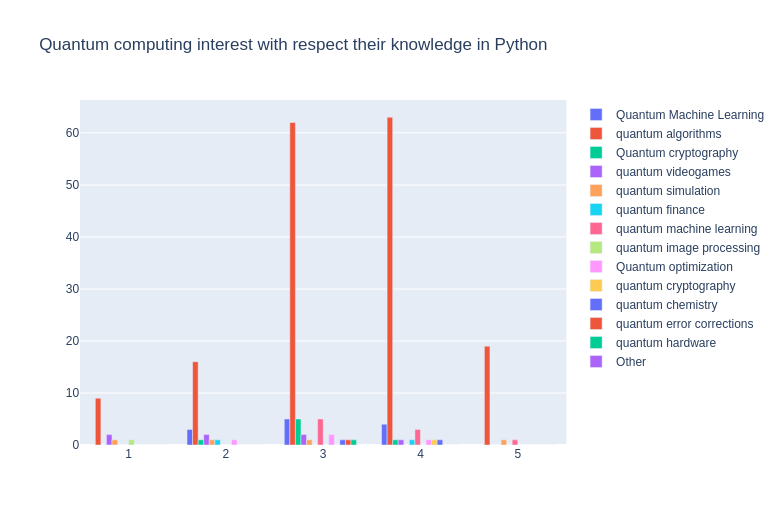}}
\subcaption{Participants' knowledge of Python compared to their interest in different areas of quantum computing.}

\label{area_python}
\end{minipage}

\begin{minipage}[c]{9cm}
\centerline{\includegraphics[width=9cm]{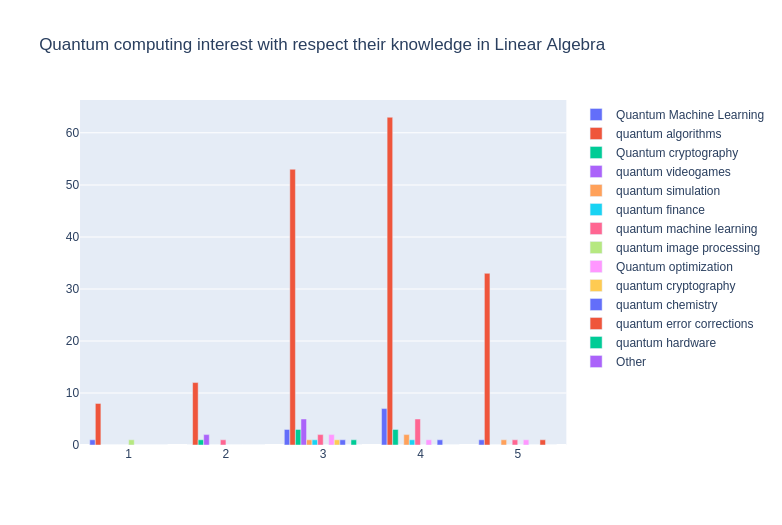}}
\subcaption{Participants' knowledge of Linear Algebra compared to their interest in different areas of quantum computing.}

\label{area_linear}
\end{minipage}
\caption{Plots of participants' self-reported interest in different areas of quantum computing, with respect to participants' self-reported knowledge of (a) quantum computing, (b) Python, and (c) linear algebra, from 1 (little to none) to 5 (expert).}
\label{knowledge}
\end{figure}

Comparison of the data on self-reported level of quantum, linear algebra, and Python competence establishes that participants of this event are generally more versed in linear algebra and Python, whilst being new to quantum computing.  More participants identified as level 1 out of 5, the lowest level, in quantum computing, than any other level.  While programming skills are expected of persons interested in a hackathon, the data on linear algebra being very similar to that is interesting.  This supports the perception of both linear algebra and programming as prerequisite skills to learning quantum, as participants felt competent in these two skills whilst being beginners in quantum (see Figure \ref{knowledge}).

It is worth noting that the participants are mostly concentrated at an introductory level in quantum computing, and most of them consider that they have intermediate to advanced knowledge about Python and linear algebra.  This is in line with the communication about the event, where it was emphasized that beginners to quantum were very welcome, and it was recommended to have some familiarity with Python and linear algebra for the workshops.

Across all participants, the subject area the most participants were interested in was by far quantum algorithms, following by quantum machine learning and quantum cryptography. This finding is interesting because it shows correspondence with a survey of 57 companies in the quantum industry, which identified quantum algorithm development as the skill most relevant across job roles in quantum computing~\cite{b18}.

\subsection{Feedback}
We collected feedback from 16 participants a week after the hackathon.  First, we note several comments regarding the data collection and storage to bear in mind when we run this workshop series and hackathon again.  At the time, the objective of soliciting feedback was to gauge from a small sample of participants the general perception of the event, in addition to identifying any areas for improvement.  In retrospect, it would have better informed future similar initiatives to collect feedback promptly at two points: at the conclusion of the week of workshops, and at the conclusion of the hackathon.  By surveying participants a week after the hackathon, we were less likely to hear from participants of only the workshops part of the event.  Another complications we had was inexperience with regards to data privacy laws to storing participant identifying information and thus opting to be on the safe side of not collecting sensitive information.  For this reason, we did not have sufficient data to draw conclusions about changes before and after the hackathon, for instance in participants' self-reported quantum computing level.  In the future, we would also like to know participants' field(s) of study for students undergraduate level and above, and aspiring topic(s) for high school students.

We report three observations pertaining to the participant feedback data.  The first is the favorite workshop(s) of each surveyed participant, presented in Table~\ref{tab3} where each row lists the favorite workshop(s) of a particular participant, sorted from lowest (top of the table) to highest (bottom) by self-reported quantum computing level, and within the same level sorted by education level.  As expected, the Introduction to quantum computing workshop was frequently favorited in the top half of the table but not in the bottom half.  The most popular topics within the workshops were QML, QAOA, and Quantum Enhanced Monte Carlo simultation.

\begin{table}[htbp]
\caption{Favorite workshop(s) of surveyed participants}
\begin{center}
\begin{tabular}{|p{5.5cm}|}
\hline
Knapsack Problem\\ \hline
Introduction to QML\\ \hline
Introduction to QML\\
Introduction to quantum computing\\
BB84 protocol\\ \hline
Introduction to quantum computing\\ \hline
Knapsack Problem\\
Introduction to QML\\
Introduction to QAOA\\ \hline
Introduction to quantum computing\\
Quantum Enhanced Monte Carlo Simulations\\
Status of quantum computing in Latin America\\
Computing physical observables with VQE\\ \hline
Quantum Enhanced Monte Carlo Simulations\\ \hline
Introduction to quantum computing\\ \hline
Introdución al QAOA\\ Introduction to QML\\
Knapsack Problem\\ \hline
Introduction to quantum computing\\
Status of quantum computing in Latin America\\
Computing physical observables with VQE\\
Knapsack Problem\\
Quantum Enhanced Monte Carlo Simulations\\ \hline
Introduction to QML\\ \hline
Knapsack Problem\\
Quantum Enhanced Monte Carlo Simulations\\ \hline
Introdution to QAOA\\
Quantum Enhanced Monte Carlo Simulations\\ \hline
Introduction to QAOA\\
Quantum Enhanced Monte Carlo Simulations\\
quantum algorithms with oracles\\ \hline
Introduction to QML\\ \hline
Quantum Enhanced Monte Carlo Simulations\\
Introduction to QAOA\\ \hline
\end{tabular}
\label{tab3}
\end{center}
\end{table}

The second observation is the highly positive perception amongst participants about the event organization.  10 of the 16 participants answered the optional short reply question to solicit feedback.  In spite of the question being about feedback, there were only two constructive criticisms received: that the Quantum Machine Learning workshop was too technical, and the online meeting links could be sent more in advance.
All the responses were in Spanish.  Their machine translation to English are provided in Table~\ref{tab4}.  Like Table~\ref{tab3}, Table~\ref{tab4} is sorted from lowest (top of the table) to highest (bottom) by self-reported quantum computing level, and within the same level sorted by education level.  We hope that by sharing their expressed wishes to see more workshops and events in Spanish, it can be seen that there is avid interest in learning quantum computing in Mexico and Latin America.

\begin{table}[htbp]
\caption{Participant optional feedback to the organizers}
\begin{center}
\begin{tabular}{|p{7.5cm}|}
\hline
I wish there were more workshops and the resources were more open and available.\\ \hline
Very interesting the topics presented.\\ \hline
Everything was excellent, I really noticed the effort of the organizers. I wish there were more events like this one. \\ \hline
I was interested in the Quantum Machine Learning workshop, but it was very technical and I got lost.\\ \hline
They were days of great learning in the quantum computing workshops, great organization of the event, I am very happy that I decided to propose my own initiative in the Hackathon, I want to live the experience and what better than in my language (Spanish) is really a plus that motivates me even more to continue learning. Thank you very much!\\ \hline
Very good project, it is a breakthrough for quantum computing in the country.\\ \hline
Nothing more. Thanks to the organizers, I appreciate the order that they had with everything and their effort is noticeable!\\ \hline
I was delighted that these workshops were given in Spanish, it facilitates the dissemination of quantum computing in Latin America.\\ \hline
You were very good, the only recommendation might be that the link was a little earlier because in some meetings it was 10 minutes earlier.\\ \hline
I think this is an excellent initiative. Keep it up :D\\ \hline
\end{tabular}
\label{tab4}
\end{center}
\end{table}

\section{Conclusion}

Science communication and outreach was done after the event, not only by the Qiskit Blog~\cite{b19}, but also outside the field of quantum reaching non-technical and Spanish speaking audiences.
In the aftermath of the event, the main event organizer was interviewed in Spanish by a national television channel in Mexico to talk about the event, and the CORE International Congress published a magazine article in Spanish about the event~\cite{b20}.
Finally, the host institute IPN conducted an interview in Spanish with the local participants who won third place~\cite{b21}.

In addition to the benefits of raising literacy in quantum technologies, and encouraging students to interact with industry and researchers, these events motivate participants to gain confidence in further pursuits learning quantum.  Furthermore, such events can connect participants with more opportunities.  For example, at the hackathon, one of the first place winners was encouraged to apply to the Qiskit advocates program by IBM Quantum, which he has since become a part of after passing the Qiskit developer certification exam.

We conclude by stating that in the participant feedback survey, for the question, ``Would you like to see more events like this in Spanish?'', $100\%$ of the responses were ``Yes''.  There is much interest in learning quantum computing and gaining relevant skills from Latin America, and lessons were learned in how to develop opportunities to catalyze this.

\section*{Acknowledgment}
This event would not have been possible without the active involvement and support of many people.  We would like to thank our fellow hackathon organizers, Jose Navarro and David P\'erez, along with CIC-IPN for their collaboration as the host organization.  We would also like to thank all of the workshop presenters in Table~\ref{tab1} and hackathon mentors.  We thank IBM Quantum, Multiverse Computing, Xanadu, QMexico, QuantumHispano, Haq.ai, and Quantum Universal Education for sponsoring this event.  We thank IBM Quantum Education for providing resources and guidance in organizing the hackathon as part of Qiskit Fall Fest 2021, especially Brian Ingmanson, Anamar\'ia Garc\'ia Hern\'andez, Katie Pizzolato, and Josie Kies.
Finally, we thank the workshop series and hackathon participants for their feedback, and for their enthusiasm which made the event a joy to organize.

\end{document}